\documentstyle[11pt,IAUS212,twoside,epsf]{article}

\def\edcomment#1{\iffalse\marginpar{\raggedright\sl#1\/}\else\relax\fi}
\reversemarginpar

\begin{document}

\title{The Massive Wolf-Rayet Binary HDE~318016~(=WR~98)}
 \author{Roberto C. Gamen\altaffilmark{1} and 
Virpi S. Niemela\altaffilmark{2} }
\affil{Facultad de Ciencias Astron\'omicas y Geof\'\i sicas, Universidad 
Nacional de La Plata.
Paseo del Bosque s/n, B1900FWA, La Plata, Argentina.}
\altaffiltext{1}{ Fellow of CONICET, Argentina}
\altaffiltext{2}{Visiting Astronomer, CTIO, NOAO, operated by AURA, Inc. 
for NSF, USA. Member of Carrera del Investigador, CIC--BA, Argentina}

\begin{abstract}
We present the discovery of OB type absorption lines superimposed to the
emission line spectrum, and the first double-lined orbital elements for the
massive Wolf-Rayet binary HDE~318016 (=WR~98), a spectroscopic binary in
a circular orbit with a period of 47.825 days. The semiamplitudes of
the orbital motion of the emission lines differ from line to line,
indicating mass ratios between 1 and 1.7 for
$\mathcal{M}_{WR}/ \mathcal{M}_{OB}$.
\end{abstract}

\section{Introduction and Observations}
HDE~318016
=WR~98 is one of the few stars with Wolf-Rayet spectrum showing both N and C 
emission lines. This star was found to be a single-lined 
binary with a period of
47.8 days with N and C emission lines moving in phase (Niemela 1991).
Relevant parameters of WR~98 can be found in the recent VIIth Catalogue
of galactic WR stars (van der Hucht  2001).

Here we present the results of a detailed radial velocity analysis of
optical spectral lines of WR~98 showing it to be a double-lined binary.

We have obtained a total of 69 blue optical spectrograms of WR~98 at two
different Observatories, namely CTIO in Chile (1980-84) and CASLEO 
(1997--2001) in Argentina.
The observations at CTIO were performed with the Cassegrain IT spectrograph 
on the 1m Yale telescope, using photographic plates with fine grain emulsion
as detector.
The observations at CASLEO where obtained with the Cassegrain 
 spectrographs with CCD detectors attached to the 2.15m telescope.
({\footnotesize CASLEO is operated under agreement between CONICET, SeCyT, 
and the Universities of La Plata, C\'ordoba and San Juan, Argentina.})

The photographic observations were digitized with a Grant engine at La Plata.
The digital data were processed and measured with IRAF routines.

\section{Results}
\begin{figure}
\plotfiddle{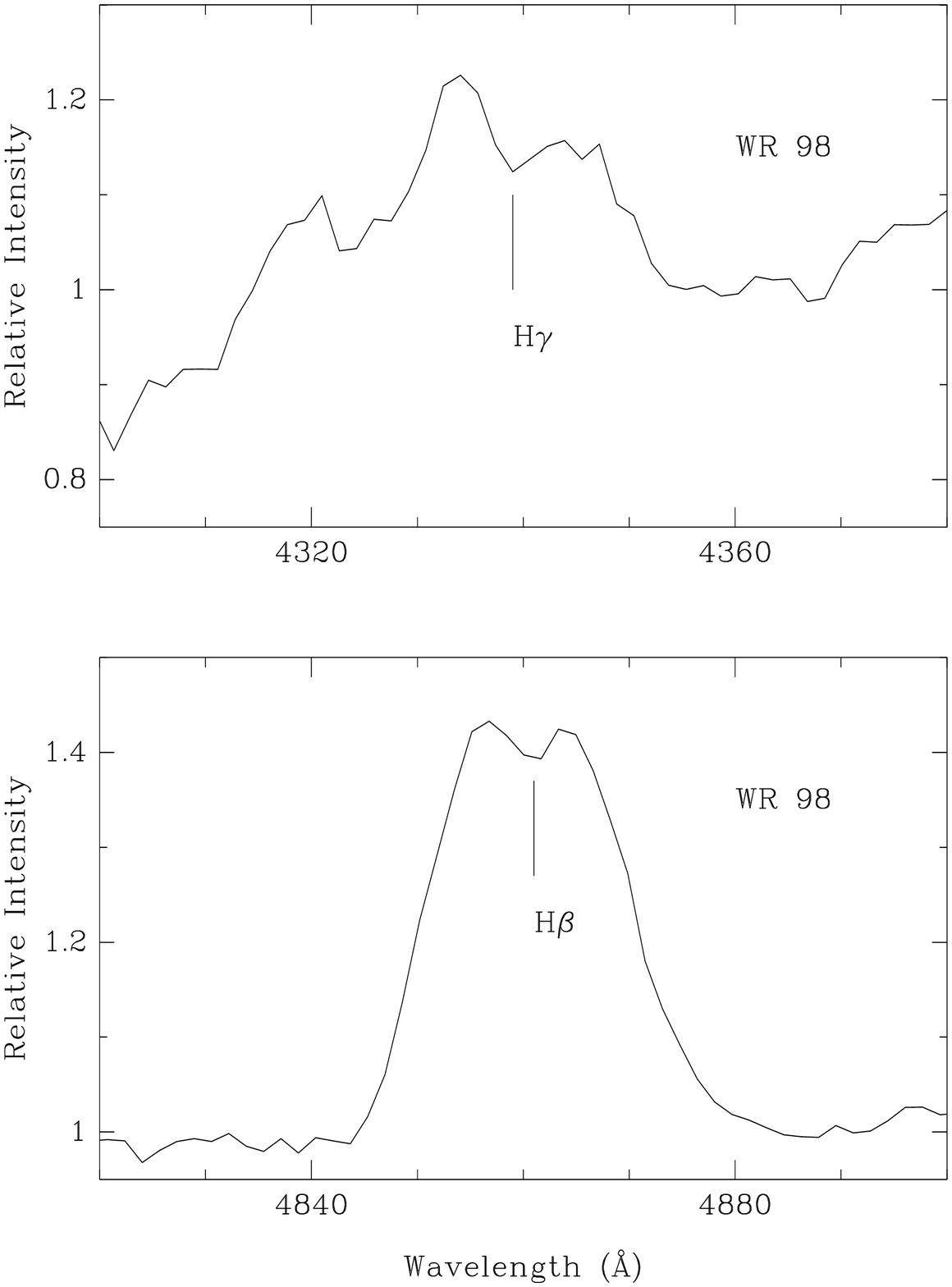}{7cm}{0}{34}{33}{-199}{-45}
\plotfiddle{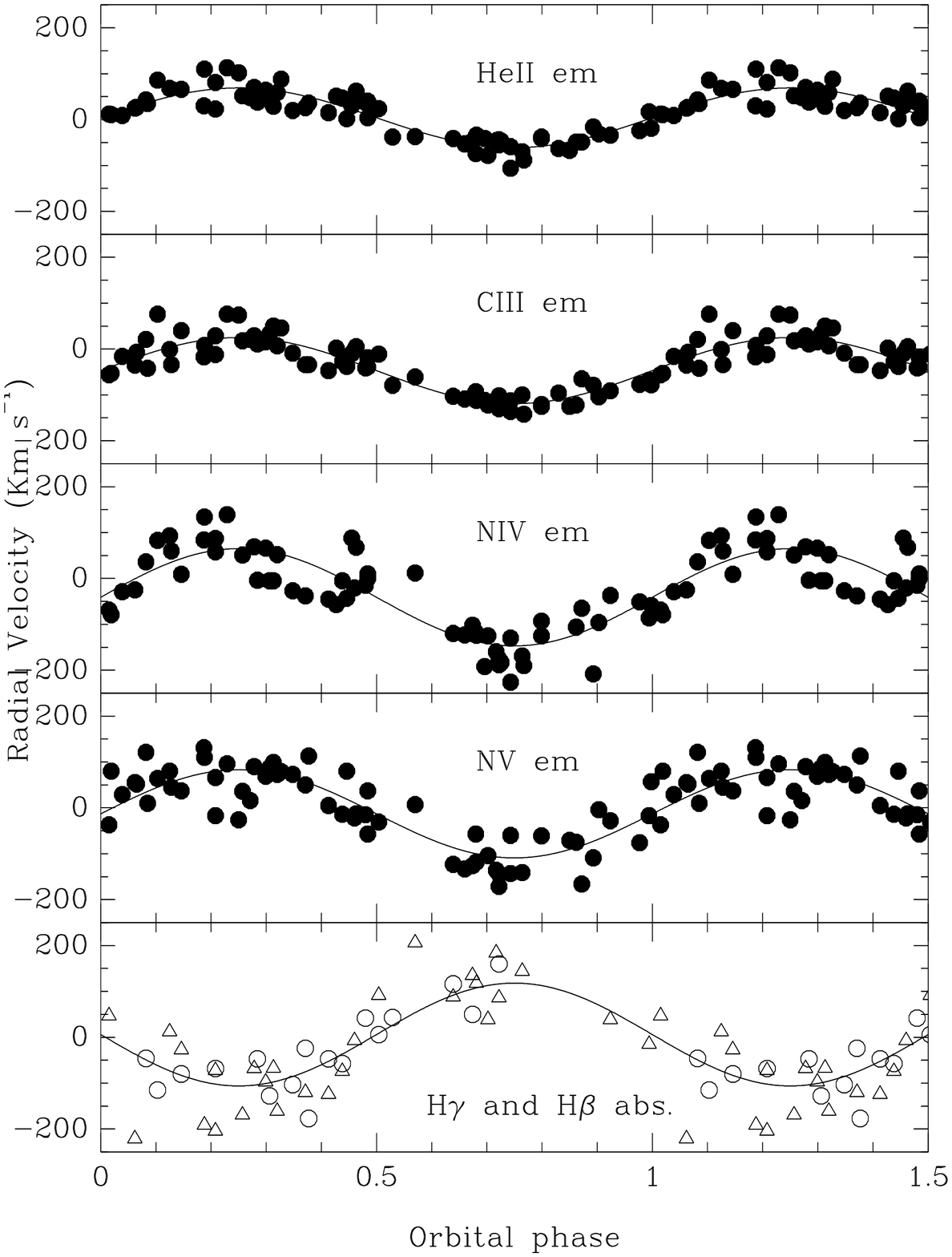}{0cm}{0}{35}{33}{-10}{-20}
\caption{Left: H$\gamma$ and H$\beta$ absorptions observed atop of the 
He{\sc ii} emissions.
Right: Radial velocity variations of emission lines, and absorption lines
of H$\gamma$ (circles) and H$\beta$ (triangles) phased with P=47.825~days.
Curves represent the orbital solutions from Table~1.}
\end{figure}

Our radial velocity analysis of the optical spectral lines of WR~98
using a long term database, confirms that this star is a binary system
in a circular orbit with a period of 47.825~days. Figure~1 depicts the
radial velocity variations of the spectral lines in this period. 
N and C emission lines
in the spectrum move in phase, indicating that they are formed in the same
stellar envelope (See Fig.~1).

We have also detected absorption lines upon the emissions.
These absorptions correspond to a late O-type spectrum, and they move
anti-phased with respect to the emission lines (see Fig.~1).
Thus  WR~98 is a double-lined spectroscopic binary. Orbital elements are
listed in Table~1.
Radial velocity variations of absorption and emission lines
indicate that the mass of the WR component is similar or higher than that of the
OB component (See Table~1).

\begin{table}
\setlength{\tabcolsep}{0.5mm}
\leavevmode
\caption[]{\small Circular Orbital Elements of WR~98 (P=47.825~days).}
\scriptsize
\begin{tabular}{rl  c rcl c rcl c rcl c rcl c rcl}
\noalign{\smallskip}\hline\noalign{\smallskip}
\multicolumn{2}{c}{Element}              &~~~~~
& \multicolumn{3}{c}{~~N\,{\sc  iv} em.~~~~~}&
& \multicolumn{3}{c}{~~N\,{\sc   v} em.~~~~~}&
& \multicolumn{3}{c}{~~C\,{\sc iii} em.~~~~~}&
& \multicolumn{3}{c}{He\,{\sc  ii} em.~~~~~~}&
& \multicolumn{3}{c}{~absorptions~~~~} \\
\noalign{\smallskip} \hline \noalign{\smallskip}
$V_{0}$ &  [km\,s$^{-1}$] &  & -41 & $\pm$ & 4 &  & -15 & $\pm$ & 3 &  & -47 & $\pm$ & 2 &  & 4 & $\pm$ & 2 &  & 6 & $\pm$ & 15 \\
$K$ &  [km\,s$^{-1}$] &  & 106 & $\pm$ & 6 &  & 109 & $\pm$ & 5 &  & 72 & $\pm$ & 3 &  & 65 & $\pm$ & 3 &  & 112 & $\pm$ & 16 \\
$\mathcal{M}_{WN}$ sin$^{3}i$ & $[\mathcal{M}_\odot]$ &  & 27 & $\pm$ & 10 &  & 28 & $\pm$ & 10 &  & 19 & $\pm$ & 8 &  & 18 & $\pm$ & 7 &  &  &   &  \\
$\mathcal{M}_{OB}$ sin$^{3}i$ & $[\mathcal{M}_\odot]$ &  & 25 & $\pm$ & 7 &  & 27 & $\pm$ & 7 &  & 12 & $\pm$ & 3 &  & 10 & $\pm$ & 3 &  &  &   &  \\
\noalign{\smallskip}
\hline
\end{tabular}
\end{table}


\end{document}